\documentclass[twocolumn,pra,showpacs,preprintnumbers,superscriptaddress]{revtex4}
\usepackage{graphicx}
\usepackage{dcolumn}
\usepackage{bm}
\usepackage{amssymb}
\usepackage{amsmath}
\usepackage{extarrows}

\def \TT {\textsf{T}}
\def \CC {\textsf{C}}
\def \PP {\textsf{P}}
\def \AA {\textsf{A}}

\begin{document}
\title{Quantum Simulation of Exotic $\textsf{P}\textsf{T}$-invariant Topological Nodal Loop Bands  with Ultracold Atoms in an Optical Lattice}

\author{Dan-Wei Zhang}
\email{zdanwei@126.com}
\affiliation{Guangdong Provincial Key Laboratory of Quantum Engineering and Quantum Materials,
SPTE, South China Normal University, Guangzhou 510006, China}
\affiliation{Department of Physics and Center of Theoretical and Computational Physics, The University of Hong Kong, Pokfulam Road, Hong Kong, China}

\author{Y. X. Zhao}
\email{yuxinphy@hku.hk}
\affiliation{Max-Planck-Institute for Solid State Research, D-70569 Stuttgart, Germany}
\affiliation{Department of Physics and Center of Theoretical and Computational Physics, The University of Hong Kong, Pokfulam Road, Hong Kong, China}

\author{Rui-Bin Liu}
\affiliation{Guangdong Provincial Key Laboratory of Quantum Engineering and Quantum Materials,
SPTE, South China Normal University, Guangzhou 510006, China}

\author{Zheng-Yuan Xue}
\affiliation{Guangdong Provincial Key Laboratory of Quantum Engineering and Quantum Materials,
SPTE, South China Normal University, Guangzhou 510006, China}

\author{Shi-Liang Zhu}
\email{slzhu@nju.edu.cn} \affiliation{National Laboratory of Solid
State Microstructures and School of Physics, Nanjing University,
Nanjing 210093, China} \affiliation{Synergetic Innovation Center of Quantum Information and Quantum
Physics, University of Science and Technology of China, Hefei
230026, China}

\author{Z. D. Wang}
\email{zwang@hku.hk}
\affiliation{Department of Physics and Center of Theoretical and Computational Physics, The University of Hong Kong, Pokfulam Road, Hong Kong, China}

\begin{abstract}
 Since the well-known $\PP\TT$ symmetry  has its fundamental significance and implication in physics, where $\PP\TT$ denotes a joint operation of space-inversion $\PP$ and time-reversal $\TT$, it is important and intriguing to explore exotic $\PP\TT$-invariant topological metals and to physically realize them. Here we
 develop a theory for a new type of topological metals that are described by a two-band model of PT-invariant topological nodal loop  states in a three-dimensional Brillouin zone, with the
 topological stability being revealed through the $\PP\TT$-symmetry-protected nontrivial $\mathbb{Z}_2$ topological charge even in the absence of both $\PP$ and $\TT$ symmetries. Moreover, the gapless boundary modes are demonstrated to be originated from the nontrivial topological charge of the bulk nodal loop.   Based on these exact results, we propose an experimental scheme to realize and to detect tunable $\PP\TT$-invariant topological nodal loop states with ultracold atoms in an optical lattice, in which atoms with two hyperfine spin states are loaded in a spin-dependent three-dimensional optical lattice and two pairs of Raman lasers are used to create out-of-plane spin-flip hopping with site-dependent phase. It is shown that such a realistic cold-atom setup can yield topological nodal loop states, having a tunable  band-touching ring with the two-fold degeneracy in the bulk spectrum and non-trivial surface states. The nodal loop states are actually protected by the combined $\PP\TT$ symmetry and are characterized by a $\mathbb{Z}_2$-type invariant (or topological charge), i.e., a quantized Berry phase. Remarkably, we demonstrate with numerical simulations that (i) the characteristic nodal ring can be detected by measuring the atomic transfer fractions in a Bloch-Zener oscillation; (ii) the topological invariant may be measured based on the time-of-flight imaging; and (iii) the surface states may be probed through Bragg spectroscopy. The present proposal for realizing topological nodal loop states in cold atom systems may provide a unique experimental platform for exploring exotic $\PP\TT$-invariant topological physics.
\end{abstract}

\date{\today}


\maketitle

\section{Introduction}

Since the discovery of topological insulators, the study of band topology of insulating and semimetallic materials has attracted
a broad interest~\cite{TI-RMP1,TI-RMP2,Volovik-book,TP-Class1,TP-Class2,TP-Class3,Kitaev,Furusaki,TP-Class4,TP-Class5,TP-Class6,Z2-Z,GaugeRMP,GaugeRPP,SOC-Review,TopoPho}.
A significant theoretical advance has been made for classification of various kinds of gapped and gapless topological band systems~\cite{Volovik-book,TP-Class1,TP-Class2,TP-Class3,Kitaev,Furusaki,TP-Class4,TP-Class5,TP-Class6}.
Very recently, a greater attention of exploring symmetry protected topological phases seems to move from gapped insulators/superconductors
to gapless metals/semimetals. For three-dimensional systems, two kinds of topological semimetals, which respectively consist of the nodal point and nodal
loop (NL), have been addressed \cite{DiracTheo,WeylTheo1,WeylTheo2,WeylTheo3,NodalTheo1,NodalTheo2}. Weyl semimetals \cite{WeylTheo1,WeylTheo2,WeylTheo3}
and certain Dirac semimetals \cite{TP-Class4,DiracTheo} belong to the former ones, which have the two-fold and four-fold degenerate Fermi points, respectively,
which are topologically protected by a topological invariant (such as the Chern number for Weyl semimetals) and give rise to exotic
Fermi-arc surface states 
\cite{DiracTheo,WeylTheo1,WeylTheo2,WeylTheo3}. While an NL
semimetal has the valence and conduction bands crossing along
closed lines instead of isolated points, which may topologically be
protected by certain discrete symmetry and may give rise to nearly
flat surface bands~\cite{NodalTheo1,NodalTheo2}. Weyl and Dirac
semimetals have been theoretically and experimentally explored not
only in
materials~\cite{TP-Class4,DiracTheo,WeylTheo1,WeylTheo2,WeylTheo3,SMExp1,SMExp2,SMExp3,SMExp4,SMExp5,SMExp7},
but also in some artificial systems, such as photonic and acoustic
crystals \cite{Lu2013,Lu2015,Xiao2015}. However, topological NL
bands are yet to be experimentally observed or realized, even
though some theoretical proposals have been suggested for their
realization in real materials very
recently~\cite{NLProp1,NLProp2,NLProp3,NLProp4,NLProp5,NLProp6}.
In addition, the so-called $\PP\TT$ symmetry may actually be understood as a
generalized  inversion symmetry with regard to the space-time
dimension, which is fundamental  in physics, and thus a realistic model 
for describing exotic $\PP\TT$-invariant
topological NLs is highly desirable. Notably, a purely combined $\PP\TT$-invariant
topological NL state that has neither $\PP$ symmetry nor $\TT$ one, can hardly be realized in real condensed materials, and thus
its physical implementation presents a great challenge to physicists both theoretically and experimentally.

On the other hand, fortunately, ultracold atoms in optical lattices
\cite{ColdAtom1} with synthetic electromagnetic field and
spin-orbit  coupling \cite{GaugeRMP,GaugeRPP,SOC-Review} provide a
clean and tunable platform for exploring exotic topological
quantum phases
\cite{Miyake,Bloch2013a,Bloch2013b,Jotzu,Bloch2015,Duca,Cold-Edge1,Cold-Edge2}.
Remarkably, the Zak phase in topologically nontrivial Bloch bands
realized in one-dimensional double-well optical lattices has been
measured \cite{Bloch2013b}. By engineering the atomic hopping
configurations in optical lattices, the celebrated
Harper-Hofstadter model \cite{HHModel} and Haldane model
\cite{HaldaneModel} have been realized
\cite{Miyake,Bloch2013a,Jotzu,Bloch2015}, where the Chern number
characterizing the topological bands has also been measured. The
experimental observation of chiral edge states in one-dimensional
optical lattices subjected to a synthetic magnetic field and an
artificial dimension has also been reported
\cite{Cold-Edge1,Cold-Edge2}. An important question then is
whether the other predicted topological phases that are rare in
sold-state materials can be realized in these cold atom systems.
Several schemes have been proposed to realize $\mathbb{Z}_2$
topological insulators \cite{Goldman2010,Copper2011}, chiral
topological states \cite{Liu2013,Liu2014,Duan2014}, and
topological nodal point semimatals
\cite{Sun2012,Jiang2012,Dubcek2015,He2015,ZDW2015,Ganeshan2015}
using ultracold atoms in optical lattices. However, an experimentally feasible
and tunable scheme for realizing the combined $\PP\TT$ topological NL states
and their detection is still badly awaited.

In this article, we first depict  the
classification of topological NLs in systems with  the pure
$\PP\TT$  symmetry. In three-dimensional momentum space, only
$\PP\TT$-invariant NLs are topologically protected with a
$\mathbb{Z}_2$ charge. We then construct a theoretical model to realize the topological nontrivial NL, and demonstrate its topological stability under $\PP\TT$ invariant perturbations that break both $\TT$ and $\PP$ symmetries. It is also shown that the nontrivial topological charge of the bulk NL directly determines the certain boundary gapless modes.  Based on these theoretical results, we propose
an experimental scheme to realize and to detect tunable
$\PP\TT$-invariant topological NL states with ultracold atoms in
an optical lattice. In our proposal, fermionic (or bosonic) atoms
with two hyperfine spin states are loaded in a spin-dependent
three-dimensional optical lattice, and two pairs of Raman lasers
are used to create out-of-plane spin-flip hopping with
site-dependent phase. We show that such a realistic cold-atom
setup can yield topological NL states having tunable ring-shaped
band-touching lines with two-fold degeneracy in the bulk spectrum
and non-trivial surface states. The NL states are actually
topologically protected by the combined $\PP\TT$ symmetry even in
the presence of $\PP$ and $\TT$ breaking perturbations, and are
characterized by a quantized Berry phase (a $\mathbb{Z}_2$-type
invariant). Moreover, with numerical simulations, we demonstrate
that (i) the characteristic nodal ring can be detected by
measuring the atomic transfer fractions in a Bloch-Zener
oscillation; (ii) the topological invariant (charge) can be
measured based on the time-of-flight imaging; and (iii) the
surface states may be probed through Bragg spectroscopy. The
realization of the combined $\PP\TT$ topological NL states in cold atom systems
would definitely be of great importance in contributing to topological matter
research across disciplines.

The paper is organized as follows. Section II describes the topological classification of NLs in systems with the
combined $\PP\TT$ symmetry, interprets physical meanings of the $\mathbb{Z}_2$ topological invariant in a lattice model, and discusses the topological stability against perturbations preserving the $\PP\TT$ symmetry but breaking both $\PP$ and $\TT$ symmetries.  In Sec. III, we propose an experimentally feasible scheme for realizing the $\PP\TT$-invariant topological NL states with ultracold atoms in an optical lattice. We first introduce the
proposed system and lattice model, and then study the tunable NLs
and their experimental detection. In Sec. IV, we elaborate the
topological properties of the simulated NL states by calculating
the quantized Berry phase and the surface states, and also present
practical methods for their experimental detection in the cold
atom system with numerical simulations. Finally, a short
conclusion is given in Sec. V.

\section{Classification of $\PP\TT$-invariant topological metals/semimetals and the topological stability of nodal loops}

In this section, we first describe a mathematically rigorous
classification of topological metals/semimetals protected by the
$\PP\TT$ symmetry using the $KO$-theory. Then we construct a
physically realistic $\PP\TT$-invariant NL model, which has a
nontrivial $\mathbb{Z}_2$ topological charge. The stability of the
NL is investigated in detail, making concrete implications of its
nontrivial topological charge. Other physical meanings of the
topological charge are also discussed. 

\subsection{Classification of topological metals/semimetals with $\PP\TT$ symmetry}

Let us consider a non-interacting fermionic system, which  is
described by the Hamiltonian $\mathcal{H}(k)$ in the  momentum space.
For such a system, the time-reversal $\TT$
and inversion $\PP$ symmetries are represented by $\hat{T}$ and  $\hat{P}$
as, respectively
\begin{equation}
\hat{T}\mathcal{H}(k)\hat{T}^{-1}=\mathcal{H}(-k),\quad \hat{P}\mathcal{H}(k)\hat{P}^{-1}=\mathcal{H}(-k).
\end{equation}
$\hat{T}$ is anti-unitary while $\hat{P}$ is unitary,
\begin{equation}
\hat{T}i\hat{T}^{-1}=-i,\quad \hat{P}i\hat{P}^{-1}=i.
\end{equation}

We here are interested mainly in the combined symmetry
$\textsf{A}=\PP\TT$, namely only $\PP\TT$ is required to be
preserved,  while $\TT$ and $\PP$ may be broken individually.  The
operation of the anti-unitary symmetry $\textsf{A}$ is given by
\begin{equation}
\hat{A}\mathcal{H}(k)\hat{A}^{-1}=\mathcal{H}(k), \quad \hat{A}i\hat{A}^{-1}=-i. \label{Sym-A}
\end{equation}
It is clear that $\AA$ operates pointwisely in the momentum space as an anti-unitary operator, and therefore the corresponding Berry bundle generated by $\mathcal{H}(k)$ in the gapped region has a real relation on each fiber, since every state $|\alpha,k\rangle$ and its complex conjugate  $|\alpha,k\rangle^*$ are related by an unitary tranformation, namely $U_A|\alpha,k\rangle^*=|\alpha,k\rangle$. The topological classification of fiber bundles with such a pointwise real relation (and dimension high enough) is given by an abelian group in the framework of the $KO$ theory, where each group element corresponds to a topological class \cite{Karoubi-book}. The Clifford algebra may be used as a powerful tool in the computation of the $KO$ groups \cite{Clifford-modules}, as being illustrated below \cite{note0}.

 To find the topological space of the gapped Hamiltonians under
the restriction of the symmetry $\AA$, we recombine the relevant
operators as $\hat{A}$, $i\hat{A}$ and $i\mathcal{H}$, which
anti-commute with each other forming a Clifford algebra,
\begin{equation}
\{\hat{A},i\hat{A}\}=0,\quad \{\hat{A},i\mathcal{H}\}=0,\quad \{i\mathcal{H},i\hat{A}\}=0.
\end{equation}
For any gapped $k$, we adiabatically flatten $\mathcal{H}(k)$ to
be $\tilde{\mathcal{H}}(k)$, which has the unital  normalization
$\tilde{\mathcal{H}}^2(k)=1_{N}$ with $1_{N}$ being the $N\times
N$ identity matrix. It is sufficient, for topological purpose, to
consider $\tilde{\mathcal{H}}$. Thus for the gapped region of the
momentum space, we assume, without loss of generality, the
following normalizations,
\begin{equation}
(i\mathcal{H})^2=-1,\quad \hat{A}^2=(i\hat{A})^2=\pm 1,
\end{equation}
where $\hat{A}^2=(i\hat{A})^2$ comes from the anti-unitarity of
$\hat{A}$ in  Eq.(\ref{Sym-A}).

For the case of $\hat{A}^2=1$, it is found that $\hat{A}$ and
$i\hat{A}$ form the Clifford algebra $C^{0,2}$, which is extended
by the $i\mathcal{H}$  to be $C^{1,2}$. Since $C^{0,2}\subset
C^{1,2}\approx C^{0,0}\subset C^{0,1}$
\cite{Clifford-modules,Karoubi-book}, it is found that the space
of the gapped Hamiltonians $\mathcal{H}(k)$ is given by the
classifying space $R_{0}$ up to homotopy. For an NL in a
three-dimensional momentum space, we may choose a circle $S^1$ to
enclose the NL from its gapped transverse dimensions, and
$\mathcal{H}|_{S^1}(k)$ gives a map from $S^1$ to $R_0$, which is
classified by the $KO$ theory as \cite{Clifford-modules,Karoubi-book}
\begin{equation}
\widetilde{KO}(S^1)\cong\pi_1(R_0)\cong \pi_0(R_1)\cong \mathbb{Z}_2.
\end{equation}
This result implies that there exists a topological nontrivial class of NLs with $\mathbb{Z}_2$ charge, whose topological stability depends solely on the combined $\TT\PP$ symmetry, regardless of individual $\TT$ and $\PP$ symmetries.

\subsection{A two-band nodal loop model and the meanings of its topological charge}

\begin{figure}
\includegraphics[scale=0.5]{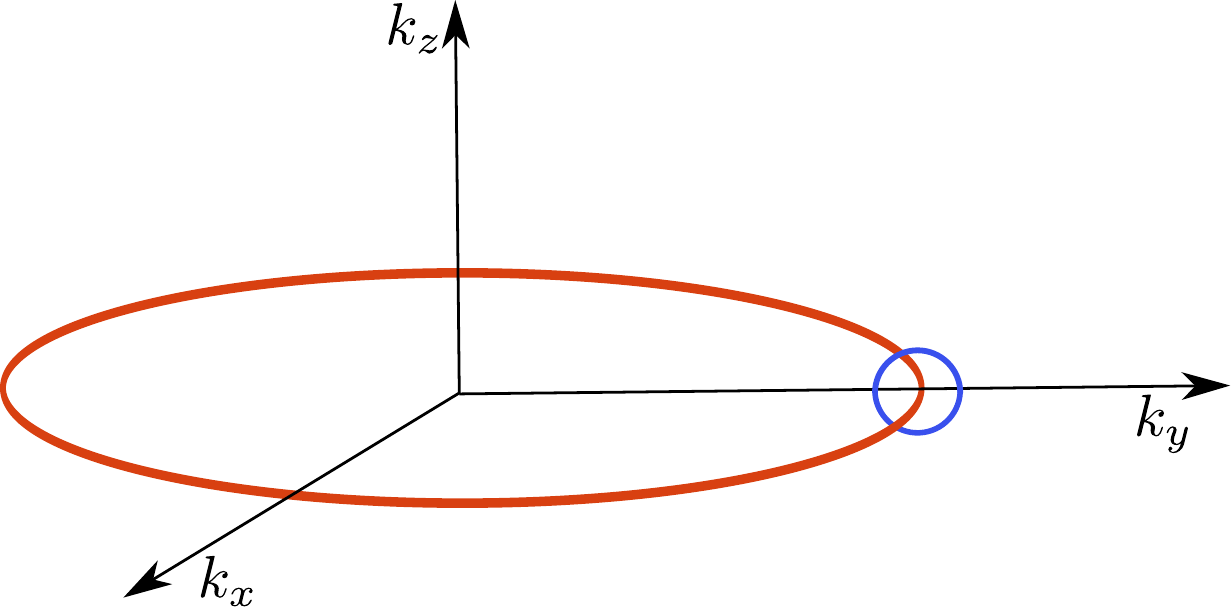}
\caption{(Color online) An NL in the three-dimensional momentum space, enclosed by a tiny circle. \label{circling}}
\end{figure}

We now construct a simple two-band model, which has $\PP\TT$
symmetry with $(\hat{P}\hat{T})^2=1$, to illustrate our theory.  A
two-band model may be written in a unified form
$\mathcal{H}(k)=f_\mu(k)\sigma^\mu$, where $f_{\mu}$ are real
functions of $k$ and $\sigma^\mu=(\sigma_0,\sigma_j)$ are Pauli
matrices. Choosing the time-reversal and inversion symmetry as
$\hat{T}=\mathcal{K}$ with the $\mathcal{K}$ as the
complex-conjugate operator and $\hat{P}=\sigma_3$, the $\PP\TT$
symmetry with $\hat{A}=\sigma_3\mathcal{K}$, requires
\begin{equation}
\sigma_3\mathcal{H}^*(k)\sigma_3=\mathcal{H}(k),
\end{equation}
which simply means the absence of $\sigma_1$ term in $\mathcal{H}(k)$. The time-reversal symmetry implies
\begin{equation}
\mathcal{H}^*(k)=\mathcal{H}(-k),
\end{equation}
leading to that the coefficients of $\sigma_0$, $\sigma_1$, and
$\sigma_3$ are even functions of $k$,  while that of $\sigma_2$ is
odd. Note that the conservation of $\TT\PP$ and $\TT$ implies also
the conservation of $\PP$. From the above points, we start with a
simple continuum model in three dimensions  as
\begin{equation}
\mathcal{H}_0(\mathbf{k})=[R^2-B(k_x^2+k_y^2)-B_z k_z^2]\sigma_3+C_z k_z\sigma_2 \label{PT-model}
\end{equation}
with $\mathbf{k}=(k_x,k_y,k_z)$ and $B>0$, which has both $\PP$
and $\TT$ symmetries, and therefore the $\PP\TT$ symmetry.  Note
that we abandon $\sigma_0$ term for clarity, since it has nothing
to do with the spectrum gap. We also keep the rotation symmetry of
the $k_x$-$k_y$ plane for simplicity. It is found that the gapless
points form an NL on the $k_x$-$k_y$ plane with $k_z=0$, which may
be enclosed by a loop from the gapped region, for instance a tiny
circle on the $k_y$-$k_z$ plane as shown in Fig. \ref{circling}.
The circle is parametrized as
$(0,R/\sqrt{B}+\rho\cos\phi,\rho\sin\phi)$, where $\rho$ is
the radius and $\phi$ the angle. If $\rho$ is sufficiently small,
the Hamiltonian restricted on the circle is expanded as
\begin{equation}
h(\phi)= -2\sqrt{B}R\rho\cos\phi\sigma_3+C_z\rho\sin\phi\sigma_2+\mathcal{O}(\rho^2).
\end{equation}
It is a well-known result that the Berry phase of the occupied state wave function of such a Hamiltonian is equivalent to one modulo $2$, namely
\begin{equation}
\gamma_{S^1}=\frac{1}{\pi}\int a(\phi) d\phi\equiv 1\mod 2,
\end{equation}
where $a(\phi)=\langle \phi|i\partial_\phi|\phi\rangle$ with
$|\phi\rangle$ being the occupied state of $h(\phi)$. The unit
topological charge $\gamma$ means the NL is in the nontrivial class
of the $\mathbb{Z}_2$ classification, which was obtained above.

According to our classification theory, the topological protection
of the stability of the gapless modes requires only the  combined
$\PP\TT$ symmetry, rather than both $\PP$ and $\TT$ symmetries. In other words, the
gapless modes still exist for topological reason under
perturbations breaking both $\PP$ and $\TT$ symmetries but
preserving $\PP\TT$ symmetry. In general, a $\sigma_2$ term with
even functions of $\mathbf{k}$ and a $\sigma_3$ term with odd
functions of $\mathbf{k}$ break both $\PP$ and $\TT$. From our
discussion of $\PP$ and $\TT$ symmetries above, the perturbations
below, for instance, satisfy the symmetry conditions.
\begin{multline}
\mathcal{H}'(\mathbf{k})=[D(k_x+k_y)+D_z k_z]\sigma_3\\
+[\mu+\epsilon_z k_z^2+\epsilon (k_x^2+k_y^2)]\sigma_2, \label{H'}
\end{multline}
under which gapless modes still survive, although the shape of the
NL has been distorted. For instance  when $\epsilon$ and
$\epsilon_{z}$ vanish, the gapless region is given by the solution
of the equation
\begin{multline}
B(k_x^2+k_y^2)-D(k_x+k_y)\\+[B_z(\mu/C_z)^2+D_z\mu/C_z-\kappa^2]=0,
\end{multline}
for which it is seen that the radius of the NL is changed and the center is moved by the perturbations.

\begin{figure}
\includegraphics[scale=0.5]{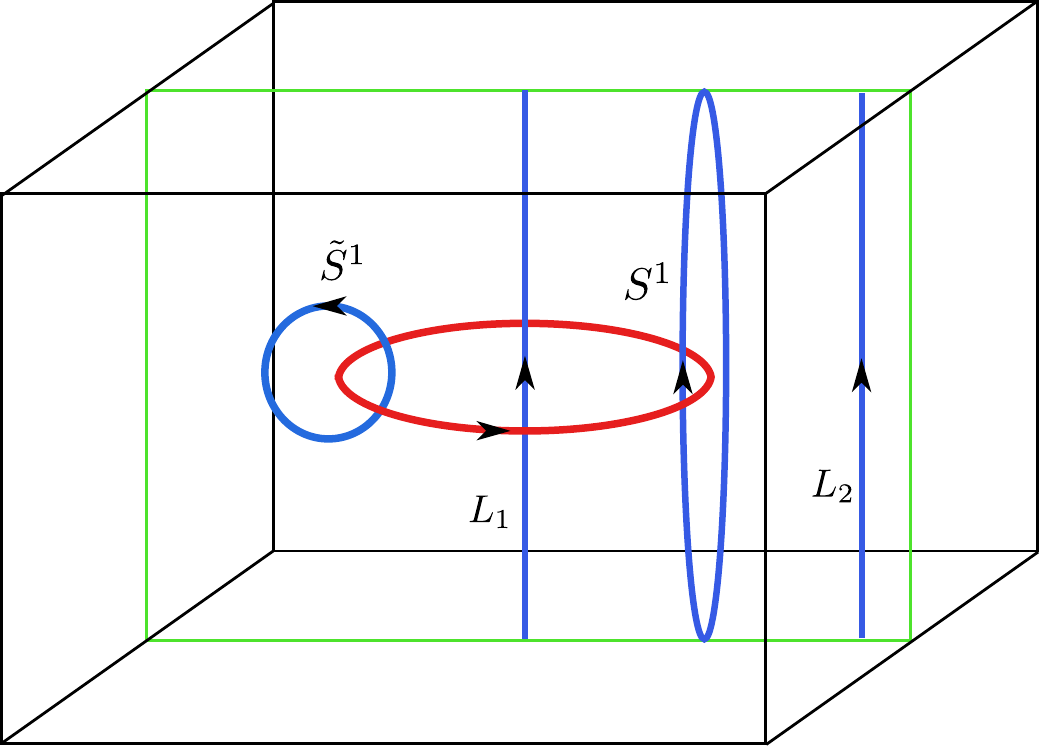}
\caption{(Color online) An NL in a three-dimensional Brillouin zone, enclosed by different loops. \label{loop-3d}}
\end{figure}

We now consider the realistic situation that the NL exists in a
lattice model, where the momentum coordinates  are periodic,
forming a Brillouin zone as illustrated in Fig. \ref{loop-3d}. The
NL in Fig. \ref{loop-3d} as the collection of gapless points is
still denoted by the red circle as in Fig. \ref{circling}. But in
a lattice model the periodicity of the momentum coordinates allows
large circles, such as $L_1$ and $L_2$ in Fig. \ref{circling},
since the two ends of such a line segment are identified. Thus for
a lattice model, the tiny circle in Fig. \ref{circling} enclosing
the NL can be continuously deformed to be the $S^1$ in the gapped
region, which may further be divided as two large circles $L_1$
and $L_2$, namely
\begin{equation}
S^1\approx L_1-L_2
\end{equation}
with the sign indicating the direction of the line. The topological charges satisfy the relation
\begin{equation}
\gamma_{S^1}=\gamma_{L_1}-\gamma_{L_2}\mod 2,
\end{equation}
where actually $L_1$ and $L_2$ can be any large circles inside and
outside the NL, respectively, as in  Fig. \ref{loop-3d}. For a
topologically nontrivial NL, if the topological charge of any
outside line is trivial, then every large circle going inside the
NL has nontrivial topological charge. For straight lines inside,
each of them may be regarded as corresponding to a one-dimensional
gapped system that is of topologically nontrivial band structure,
leading to certain gapless boundary modes.

The arguments above are applicable to general cases in the
classification. We may note an essential  difference between
$\mathbb{Z}$ and $\mathbb{Z}_2$ topological charges, namely for
$\mathbb{Z}$ topological charge, reversing the direction also
reverses the topological charge, while $\mathbb{Z}_2$ topological
charge is insensitive to the direction since $1\equiv -1\mod 2$
\cite{Z2-Z}.

\section{PT-invariant nodal loop states in cold atom systems}

Although the two-band Hamiltonian of $\PP\TT$-invariant NL is
simple, the direct implementation of this ideal model in
electronic materials is difficult due to the spin-orbit coupling
or complex lattice structure therein \cite{NLProp1,NLProp2,NLProp3,NLProp4,NLProp5,NLProp6}. In this
section, we turn to propose an experimental scheme to realize and
detect tunable $\PP\TT$-invariant topological NL states with
ultracold atoms in a three-dimensional optical lattice. We first
describe the proposed system and the lattice model, which may have both
$\PP$ and $\TT$ symmetries or generically only preserves the combined $\PP\TT$
symmetry while breaks both of them. Then we study the tunable NLs
and their experimental detection in this cold atom system.

\subsection{Proposed cold atom system and lattice model}

\begin{figure}[tbp]
\includegraphics[width=8.5cm]{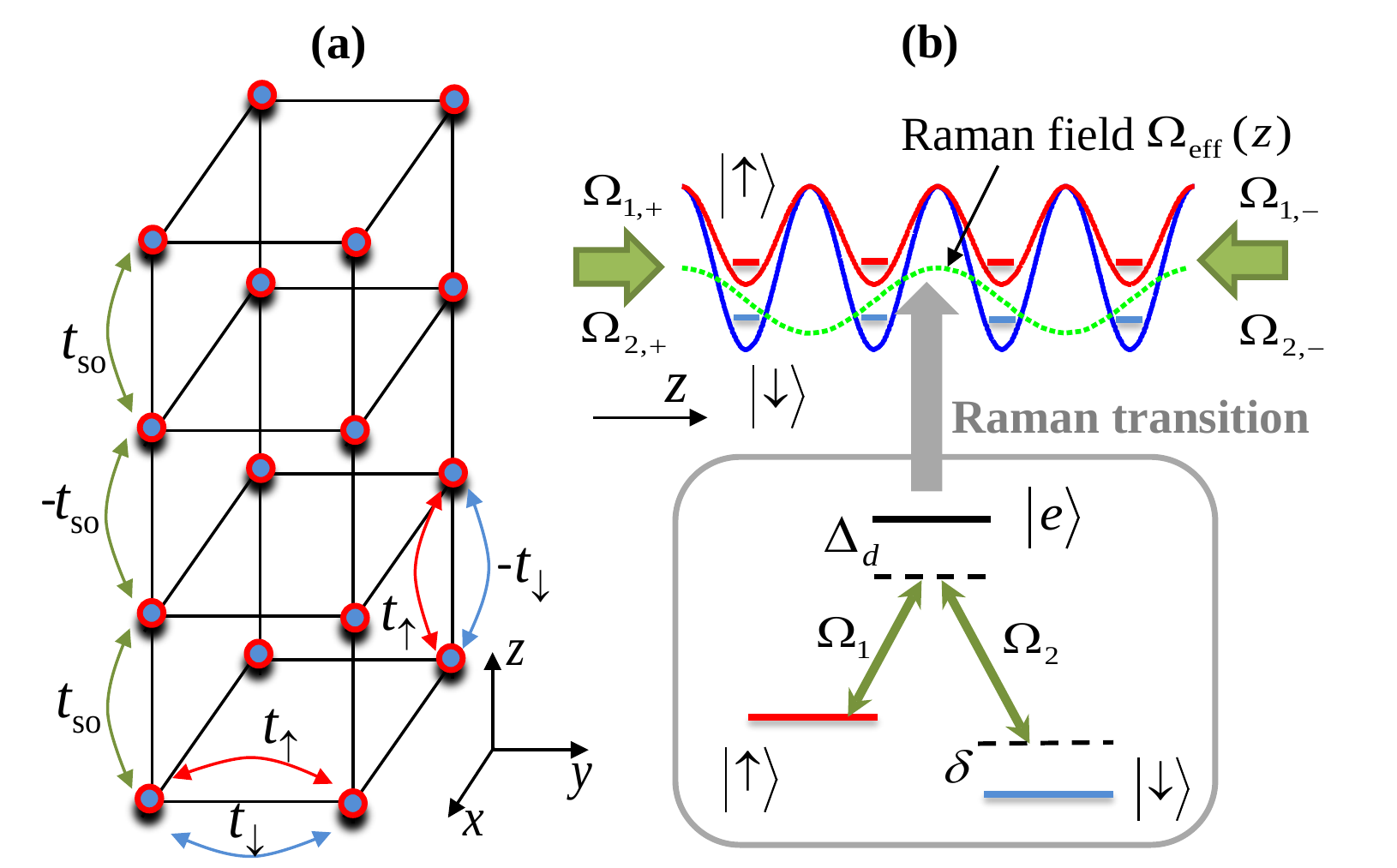}
\caption{(Color online) The proposed cold atom system and laser
configuration.  (a) The cubic optical lattice and the effective
atomic hopping configuration between nearest-neighbor lattice
sites. It contains the spin-conserved hopping $t_{\uparrow}$ along
each axis, $t_{\downarrow}$ in the $xy$ plane but
$-t_{\downarrow}$ out of the plane, the spin-flip hopping from
$\left|\downarrow\rangle\right.$ to $\left|\uparrow\rangle\right.$
which is $\pm t_{\rm so}$ staggered along the $z$ axis, and its
hermitian conjugation progress. (b) The spin-dependent lattice
structure and laser configuration of the Raman transition between
spin states $\left|\uparrow\rangle\right.$ and
$\left|\downarrow\rangle\right.$ along the $z$ axis . The
effective Raman field strength proportional to $\sin(k_0z)$ (green
dotted line) consists two pairs of plane waves denoted by their
Rabi frequencies $\Omega_{1,\pm}$ (gives $\Omega_{1}$ laser field)
and $\Omega_{2,\pm}$ (gives $\Omega_{2}$ laser field). Two spin
states are coupled via a two-photon Raman transition with a large
detuning $\Delta_d$ from an excited state $|e\rangle$ and a
two-photon detuning $\delta$. \label{system}}
\end{figure}

Our proposed system is based on a fermionic gas (or bosonic gas)
of noninteracting atoms with two  chosen hyperfine spin states
$\left|\uparrow\rangle\right.$ and
$\left|\downarrow\rangle\right.$ in a spin-dependent cubic optical
lattice, as shown in Fig. \ref{system}(a). The lattice potential
takes the form
\begin{equation}
V_{\sigma}(\mathbf{r})=-V_{0,\sigma}[\cos^2(k_0x)+\cos^2(k_0y)+\cos^2(k_0z)],
\end{equation}
where $\sigma=\uparrow,\downarrow$, $V_{0,\sigma}$ denote the
potential strengths for the spin  state $|\sigma\rangle$, and
$k_0$ is the wave number with the lattice constant $a=\pi/k_0$. As
shown in Fig. \ref{system}(a), we consider a two-photon Raman
transition between $\left|\uparrow\rangle\right.$ and
$\left|\downarrow\rangle\right.$ along the $z$ axis, which is
achieved by coupling the two ground hyperfine states to an excited
state $\left|e\rangle\right.$ with a large single-photon detuning
$\Delta_d$. In addition, the two hyperfine states differ in the
magnetic quantum number by one and thus the atomic addressing is
achieved through polarization selection. This configuration has
been used to create the equal-Rashba-Dresselhaus spin-orbit
interaction \cite{GaugeRMP,GaugeRPP,SOC-Review}. The Rabi
frequencies of the corresponding laser fields are chosen as
$\Omega_1(z)=\Omega_{0}\cos(k_0z/2)e^{i k'_0z}$ and
$\Omega_2(z)=\Omega_{0}\sin(k_0z/2)$, respectively, where
$\Omega_0$ is the strength constant controlled by the laser
intensities of the Raman fields and $k'_0$ denotes a small
deviation of the wave numbers ($k'_0\ll k_0$) that can be
tuned from zero to finite value via the laser beams. Here
$\Omega_1$ and $\Omega_2$ can be realized, respectively, with a
pair of laser beams $\Omega_{1,\pm}=\frac{1}{2}\Omega_0 e^{\pm
i(k_0\pm k'_0)z}$ and $\Omega_{2,\pm}=\frac{1}{2}\Omega_0
e^{\pm i(k_0z+\pi/2)}$. Generally, there is a two-photon detuning
$\delta$ in the Raman transition, as shown in Fig.
\ref{system}(b).

In the presence of large detuning $|\Delta|\gg\Omega_{0},|\delta|$
for the Raman transition,  we can eliminate the excited state and
consider the atomic motions in the ground state manifold. Under
this condition, the effective single-particle Hamiltonian in the
spin basis
$\{\left|\uparrow\rangle\right.,\left|\downarrow\rangle\right.\}$
is given by
\begin{equation}
H_{0}=\frac{\mathbf{p}^2}{2m_a}+\left(
                                      \begin{array}{cc}
                                        V_{\uparrow}(\mathbf{r}) & \hbar\Omega_{\text{eff}}(z) \\
                                         \hbar\Omega^*_{\text{eff}}(z) & V_{\downarrow}(\mathbf{r})+\hbar\delta \\
                                      \end{array}
                                    \right),
\end{equation}
where $\mathbf{p}$ and $m_a$ respectively denote the atomic
momentum and mass,  and
$\Omega_{\text{eff}}(z)=\frac{\Omega_1^{*}\Omega_2}{\Delta_d}=\frac{\Omega_0^2
e^{-ik'_0z}}{2\Delta_d}\sin(k_0z)$ denotes the resulting
Raman coupling in the two-photon transition. We consider all the
atoms in the lowest band of the optical lattice, and then the
Hamiltonian in the second quantization formalism takes the form
\begin{equation}
\hat{H} = \int d^3\mathbf{r}' \Psi^{\dag}(\mathbf{r}') H_0 \Psi(\mathbf{r}').
\end{equation}
Here $\Psi=(\Psi_{\uparrow},\Psi_{\downarrow})^{\text T}$ is the
two-component  field operator with
$\Psi_{\sigma}(\mathbf{r'})=\sum_{\mathbf{r}}\hat{a}_{\mathbf{r,\sigma}}w_{\sigma}(\mathbf{r'-r})$,
which is expanded in terms of the Wannier function
$w_{\sigma}(\mathbf{r'-r})$ and $\hat{a}_{\mathbf{r},\sigma}$
representing the annihilation operator of the fermionic mode of
the spin state $|\sigma\rangle$ at the lattice site $\mathbf{r}$.
A straightforward calculation yields the tight-binding Hamiltonian
\begin{eqnarray}\label{HLattice1}
&&\hat{H} =-\sum_{\mathbf{r},\sigma,\eta}t_{\sigma}\hat{a}^{\dag}_{\mathbf{r},\sigma}\hat{a}_{\mathbf{r+\hat{\eta}},\sigma}+\text{H.c.}\nonumber\\
&&~~~~~~+\sum_{\mathbf{r},\pm}t_s^{\mathbf{r,r\pm\hat{z}}}\hat{a}^{\dag}_{\mathbf{r},\uparrow}\hat{a}_{\mathbf{r\pm\hat{z}},\downarrow}+\text{H.c.}\\
&&~~~~~~+\sum_{\mathbf{r}}m_z(\hat{a}^{\dag}_{\mathbf{r},\uparrow}\hat{a}_{\mathbf{r},\uparrow}-\hat{a}^{\dag}_{\mathbf{r},\downarrow}\hat{a}_{\mathbf{r},\downarrow}).\nonumber
\end{eqnarray}
Here the spin-conserved hopping along the $\eta$ axis
($\eta=x,y,z$) is  derived as $t_{\sigma}=-\int
d^3\mathbf{r'}w^*_{\sigma}(\mathbf{r'-r})[\frac{\mathbf{p}^2}{2m_a}+V_{\sigma}]w_{\sigma}(\mathbf{r'-r-\hat{\eta}})\approx(4/\sqrt{\pi})V_{0,\sigma}^{3/4}E_R^{1/4}e^{-2\sqrt{V_{0,\sigma}/E_R}}$
in the harmonic approximation (by using Gaussian wavefunctions of the ground state centered in each lattice well for the Wannier functions)  with $E_R=\hbar^2k_0^2/2m_a$ being
the recoil energy \cite{BH}. The spin-flip hopping terms induced
by the Raman field take the form
$t_s^{\mathbf{r,r\pm\hat{z}}}=\int
d^3\mathbf{r'}w^*_{\uparrow}(\mathbf{r'-r})\hbar
\Omega_{\text{eff}}(\mathbf{r}')w_{\downarrow}(\mathbf{r'-r\mp\hat{z}})$.

In this lattice system, the strength of the Raman field
proportional to $\sin(k_0z)$ and  the lowest band Wannier
functions are antisymmetric and symmetric with respect to the
center of each lattice site, respectively. Due to this spatial
configuration, the periodic field does not couple the intra-site
spins and the spin-flip hopping terms satisfy
$t_s^{\mathbf{r,r\pm\hat{z}}}=\pm(-1)^{z/a}e^{\pm i\varphi}t_{\rm
so}$ \cite{Liu2013}, where $t_{\rm
so}=\frac{\hbar\Omega_0^2}{2\Delta_d}\int
dxw_{\uparrow}^{\ast}(x)w_{\downarrow}(x)\times\int dyw^{\ast
}_{\uparrow}(y)w_{\downarrow}(y)\times\int dzw^{\ast
}_{\uparrow}(z)e^{ik'_0z}\sin(k_0z)w_{\downarrow}(z-a)$ and
$\varphi=k'_0a$. The last term
$m_z=(V_{0,\uparrow}-V_{0,\downarrow}-\hbar\delta)/2$ is
equivalent to a Zeeman field along the $z$ axis and can be
precisely tuned via the laser frequencies of the Raman beams with
acoustic-optic modulator or through the lattice potentials for
fixed laser frequencies. By redefining the spin-down operator
$\hat a_{\mathbf{r},\downarrow}\rightarrow e^{i\pi z/a}\hat
a_{\mathbf{r},\downarrow}$, Hamiltonian (\ref{HLattice1}) can be
rewritten as
\begin{eqnarray}\label{HLattice2}
&&\hat{H} =-\sum_{\mathbf{r},\sigma}t_{\sigma}(\hat{a}^{\dag}_{\mathbf{r},\sigma}\hat{a}_{\mathbf{r+\hat{x}},\sigma}+\hat{a}^{\dag}_{\mathbf{r},\sigma}\hat{a}_{\mathbf{r+\hat{y}},\sigma})+\text{H.c.}\nonumber\\
&&~~~~~~-\sum_{\mathbf{r}}(t_{\uparrow}\hat{a}^{\dag}_{\mathbf{r},\uparrow}\hat{a}_{\mathbf{r+\hat{z}},\uparrow}-t_{\downarrow}\hat{a}^{\dag}_{\mathbf{r},\downarrow}\hat{a}_{\mathbf{r+\hat{z}},\downarrow})+\text{H.c.}\\
&&~~~~~~+\sum_{\mathbf{r}}t_{\rm so}(e^{i\varphi}\hat{a}^{\dag}_{\mathbf{r},\uparrow}\hat{a}_{\mathbf{r+\hat{z}},\downarrow}-e^{-i\varphi}\hat{a}^{\dag}_{\mathbf{r},\uparrow}\hat{a}_{\mathbf{r-\hat{z}},\downarrow})+\text{H.c.}\nonumber\\
&&~~~~~~+\sum_{\mathbf{r}}m_z(\hat{a}^{\dag}_{\mathbf{r},\uparrow}\hat{a}_{\mathbf{r},\uparrow}-\hat{a}^{\dag}_{\mathbf{r},\downarrow}\hat{a}_{\mathbf{r},\downarrow})\nonumber.
\end{eqnarray}
Figure \ref{system}(a) also shows this effective atomic hopping configuration for the $\varphi=0$ (i.e., $k'_0=0$) case in the optical lattice.

In the three-dimensional Brillouin zone, the resultant Bloch Hamiltonian is given by
\begin{equation}\label{HK}
\mathcal{H}_B = f_z(\mathbf{k})\sigma_3-2t_{\rm so}\sin(k_za+\varphi)\sigma_2-f_0(\mathbf{k})\sigma_0,
\end{equation}
where $f_z(\mathbf{k})=m_z-\alpha_{-}[\cos(k_xa)+\cos(k_ya)]-\alpha_{+}\cos(k_za)$ and $f_0(\mathbf{k})=\alpha_{+}[\cos(k_xa)+\cos(k_ya)]+\alpha_{-}\cos(k_za)$, with $\alpha_{\pm}\equiv t_{\uparrow}\pm t_{\downarrow}$ are also tunable parameters. The Bloch Hamiltonian can be rewritten as
\begin{equation}\label{HK2}
\mathcal{H}_B = \mathcal{\tilde{H}}_0+\mathcal{H}_P,
\end{equation}
where $\mathcal{\tilde{H}}_0=f_z(\mathbf{k})\sigma_3-2t_{\rm
so}\cos\varphi\sin(k_za)\sigma_2-f_0(\mathbf{k})\sigma_0$ and the
perturbation part $\mathcal{H}_P=-2t_{\rm
so}\sin\varphi\cos(k_za)\sigma_2$. Here $\mathcal{\tilde{H}}_0$
preserves both $\PP$ and $\TT$ symmetries but $\mathcal{H}_P$
(which vanishes in the case $\varphi=0$) breaks the two
symmetries. However, the whole Hamiltonian $\mathcal{H}_B$
satisfies the combined $\PP\TT$ symmetry. As analysized in Sec.
II, this guarantees the existence and the topological stability of
symmetry-protected NLs in three-dimensional Brillouin zone. The
perturbation part $\mathcal{H}_P$ in this system only shifts the
center of the NLs by replacing $k_za\rightarrow k_za+\varphi$, and
do not modified their shape and topological properties. Therefore,
without loss of generality, we take $\varphi=0$ case in the
following sections to study the NL states and their detection in
this system. Note that the proposed optical lattice system
and the Raman coupling scheme are also applicable to the bosonic
atoms \cite{Miyake,Bloch2013a,Bloch2015,GaugeRMP,GaugeRPP,SOC-Review}. As for other different types of NL states, a four-band model allowing Dirac or Weyl  rings was also recently suggested to be simulated with cold atoms \cite{XuZhang}.

\begin{figure}[tbph]
\includegraphics[width=8cm]{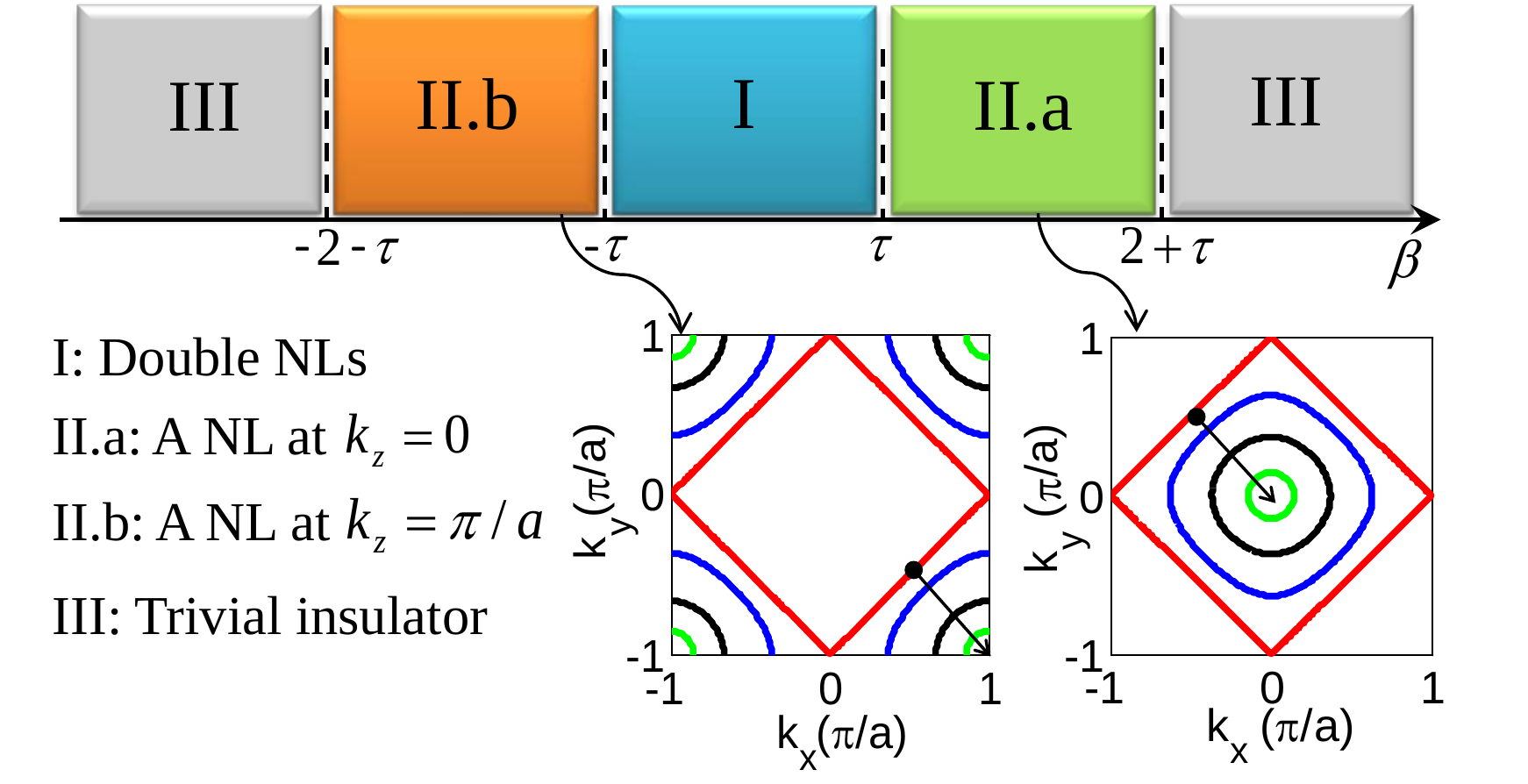}
\caption{(Color online) Phase diagram and tunable NL states. (a)
Phase diagram. III, II.a,II.b and  I correspond to trivial
insulator, an NL on the $k_z=0$ plane, an NL on the $k_z=\pi/a$
plane and two coexisting NLs on the two planes. (b) The evolution
of an NL during the increase of $\beta$ denoted by the black
arrows in (a). (c) The distribution of the topological numbers
of one-dimensional subsystems on $k_x$-$k_y$ plane in
difference phases during the increase of $\beta$ marked by the
white arrow in (a). \label{phase-diagram}}
\end{figure}

\subsection{Tunable nodal loop states}

Now we proceed to study the properties and the detection method of
the NL states in the proposed model  system. For $\varphi=0$, the
bulk spectrum is
$E_{\pm}(\mathbf{k})=f_0(\mathbf{k})\pm\sqrt{4t^2_{\rm
so}\sin^2(k_za)+f_z(\mathbf{k})^2}$. The bulk bands are fully
gapped except the points in momentum positions that satisfy the
following conditions:
\begin{eqnarray}
\cos(k_xa)+\cos(k_ya)= (m_z-\alpha_{+})/\alpha_{-}~\text{for}~k_z = 0, \label{NLFun1} \\
\cos(k_xa)+\cos(k_ya)= (m_z+\alpha_{+})/\alpha_{-} ~\text{for}~k_z = \pi/a, \label{NLFun2}
\end{eqnarray}
which can give rise to NLs with two-fold degeneracy in the
three-dimensional Brillouin zone. Without  loss of generality, we
assume $t_{\uparrow}\geq t_{\downarrow}>0$. Since the effective
Zeeman field and the hopping amplitudes can be tuned
independently, we can define two ratio parameters
$\beta=m_z/\alpha_-$ and $\tau=\alpha_+/\alpha_-$ in this system.

We then numerically solve the Eqs. (\ref{NLFun1}) and
(\ref{NLFun2}) for the existence and the shape of the  NLs, and
the resultant phase diagram is shown in Fig.
\ref{phase-diagram}(a), where the phases III, II.a, II.b and I,
respectively, correspond to trivial insulator, an NL on the
$k_z=0$ plane, an NL on the $k_z=\pi/a$ plane and two coexisting
NLs on the two planes. The two NLs belong to the same nontrivial
$\mathbb{Z}_2$ case as that being discussed in the previous
section. During the increase of $\beta$ (denoted by the black
solid arrows in Fig. \ref{phase-diagram}), the NLs have the same
evolution procession as illustrated in Fig. \ref{phase-diagram}(b).
First, a singular point is created at the corner of the
sub-Brillouin zone with $k_z=0$ or $\pi/a$, then spread to be a
circle centered at the corner. The circle expands bigger and
bigger, going across the whole sub-Brillouin zone (the red large
circle), then becomes a circle centered at the origin, and finally
disappears after decaying as a singular point at the origin. After
such a $\mathbb{Z}_2$ nontrivial small nodal circle is created at
the corner of the sub-Brillouin zone, the topological number of
any one-dimensional subsystem parametrized by $k_z$ inside the
small circle is increased by one, while any one outside has its
topological number unchanged, which can be seen from the model
discussed in Fig. \ref{loop-3d} considering a fact that the
creation of the circle results in only continuous deformations for
the outside ones. Having this in mind we can infer the topological
number of any one-dimensional system with fixed in-plane momentum $\mathbf{k}_{\|}\equiv(k_x,k_y)$ in any region of the phase
diagram. For instance, the distribution of topological number
$\gamma$ in the $k_x$-$k_y$ plane along the white arrow in Fig.
\ref{phase-diagram}(a) is shown in Fig. \ref{phase-diagram}(c),
which has also been confirmed by the numerical simulation. It is
noted that in the fifth sub-figure with $\beta=0$, ones at the
corner of the sub-Brillouin zone have the topological number
$\gamma=1+1 \equiv 0\mod 2$. In addition, the shape of the NL can
changes from a circle to square as shown in Fig.
\ref{phase-diagram}(b) due to the fact that the $k^2_{\eta}$ terms
in the continuum Hamiltonian (\ref{PT-model}) are replaced by the
$\cos(k_{\eta}a)$ terms in this lattice system with Hamiltonian
(\ref{HK}). Thus, in the proposed optical lattice system, one can
realize tunable NL states by simply varying the ratio parameter
$\beta$ via the laser fields.

\begin{figure}[tbph]
\includegraphics[width=8.5cm]{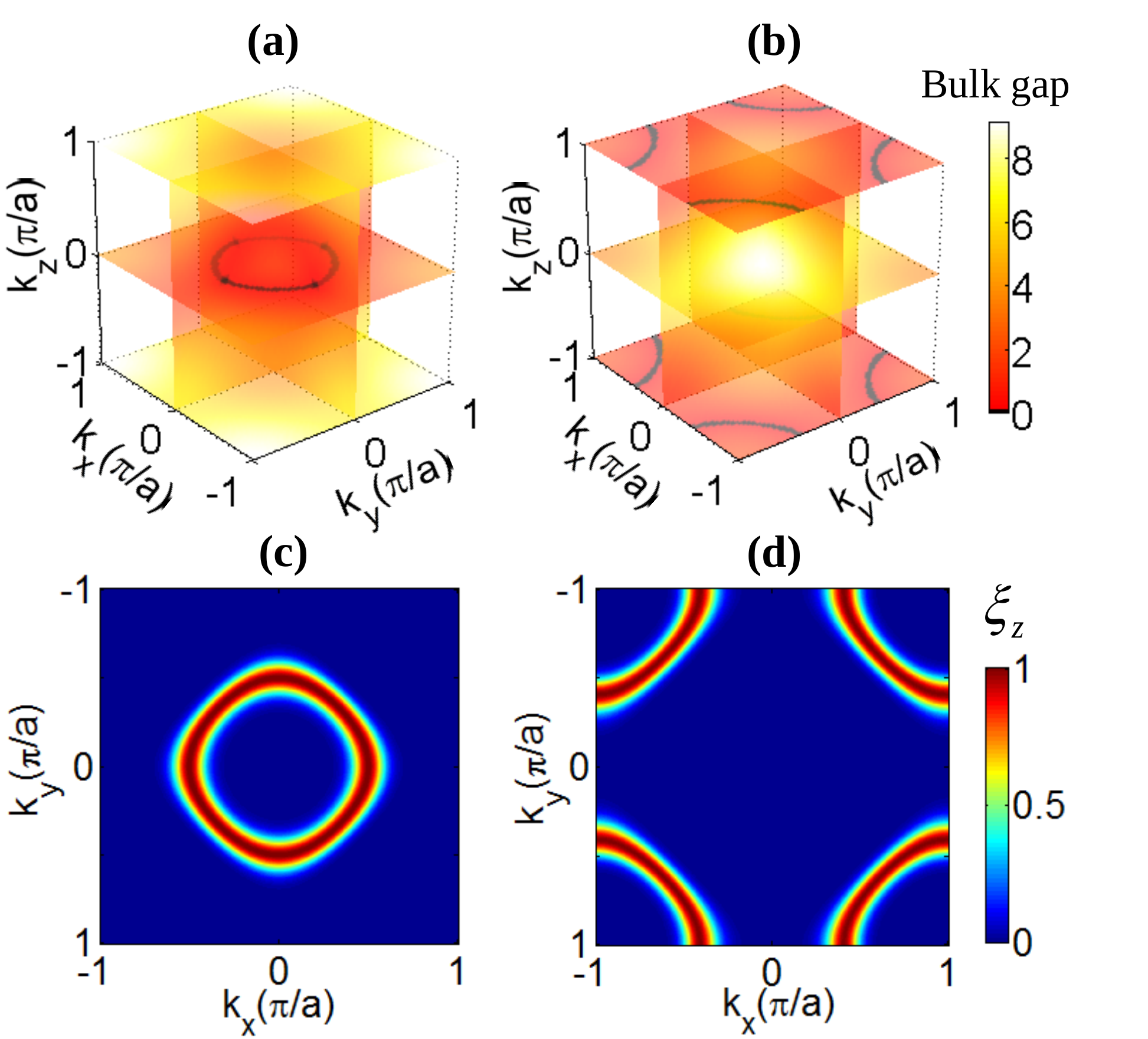}
\caption{(Color online) The bulk gap with the gap-closing points
forming a nodal ring and its detection. (a)  A nodal ring on the
$k_z=0$ plane with the parameter $m_z=2.0$. (b) A nodal ring on
the $k_z=\pi/a$ plane with the parameter $m_z=-1.8$. (c,d) The
momentum distribution of the atomic transfer fraction
$\xi_{z}(k_x,k_y)$ with the parameter $F=0.2$. The maximum
positions form a curve that reveals the nodal ring in (a) and (b),
respectively. Other parameters in (a-d) are $t_{\uparrow}=1$ as
the energy unit, $t_{\downarrow}=t_{\rm so}=0.3$.\label{probe-NL}}
\end{figure}

In the following, we focus on the single NL cases and further
study the properties of the NL state. In Figs.  \ref{probe-NL}(a)
and \ref{probe-NL}(b), we plot the bulk energy gap
$E_{+}(\mathbf{k})-E_{-}(\mathbf{k})$ for typical parameters,
which clearly show the gap-closing points forming the NLs on the
$k_z=0$ and $k_z=\pi/a$ planes, respectively. On the two planes,
the upper and lower energy surfaces touch along a ring with a
constant energy $\varepsilon_a$ in phase II.a and $\varepsilon_b$
in phase II.b. The constant energies
$\varepsilon_a=-\alpha_{+}(\alpha_{+}-m_z)/\alpha_{-}-\alpha_{-}$
and
$\varepsilon_b=-\alpha_{+}(\alpha_{+}+m_z)/\alpha_{-}+\alpha_{-}$
come from the $f_0(\mathbf{k})\sigma_0$ term in Hamiltonian
(\ref{HK}), respectively. This term breaks the chiral symmetry but
only shifts the nodal ring from zero energy without modifying its
shape. Actually, it does not affect the topological stability of
the NLs but gives the surface states a dispersion
\cite{Volovik-book,NodalTheo1}. In phase II.a, the low-energy
effective Hamiltonian near $\mathbf{K_r}=(0,0,0)$ is given by
\begin{eqnarray}\label{HEFF}
&\mathcal{H}_{\text{eff}}(\mathbf{q})= [b_{1}(q_x^2+q_y^2)+b_0q_z^2-\lambda_1]\sigma_3-v_zq_z\sigma_2 \nonumber\\
&~~~+[b_{0}(q_x^2+q_y^2)+b_1q_z^2-\lambda_0]\sigma_0,
\end{eqnarray}
where $\mathbf{q=k-K_r}$, $b_{0}=a^2\alpha_{+}/2$, $b_{1}=a^2\alpha_{-}/2$, $\lambda_1=2\alpha_{-}+\alpha_{+}-m_z$, $v_{z}=2at_{\rm so}$, and $\lambda_0=2\alpha_{+}+\alpha_{-}$. In phase II.b, the low-energy effective Hamiltonian near $\mathbf{K'_r}=(\pi/a,\pi/a,\pi/a)$ is $\mathcal{H}'_{\text{eff}}(\mathbf{q})=-\mathcal{H}_{\text{eff}}(\mathbf{q})$ with redefined $\mathbf{q=k-K'_r}$ in this case. The effective Hamiltonian (\ref{HEFF}) takes the form of the two-band model (\ref{PT-model}), and captures the essential features of the nodal ring states on the $q_z=0$ plane of radius $\sqrt{\lambda_1/b_1}$ when its size is small (see Fig. \ref{phase-diagram}).

At this stage, we turn to elaborate that the NLs can be
detected using the Bloch-Zener-oscillation method,  which has been
experimentally demonstrated to probe the Dirac points in a
honeycomb optical lattice \cite{Tarruell,Lim}. A basic idea lies
in that the band touching points can be monitored from the atomic
fraction tunnelling to the excited band in Bloch oscillations. We
can use noninteracting fermionic atoms or an incoherent
distribution of bosonic atoms with the population being
homogeneous in the momentum space initially prepared in the lower
band \cite{Bloch2015}. A constant force $F$ is applied along the
$z$ axis and pushes the atoms moving along the $k_{z}$ direction.
After a Bloch cycle, we can obtain the momentum distribution of
the transfer fraction to the upper band from time-of-flight
imaging \cite{Tarruell,Lim}. In this case, the transfer fraction
$\xi_{z}(\mathbf{k_{\|}})$ is given by
\begin{equation}\label{LZ}
\xi_{z}(\mathbf{k_{\|}})=P_{\text{LZ}}(\mathbf{k_{\|}})=e^{-\pi\Delta_z^2(\mathbf{k_{\|}})/4v_{z}F},
\end{equation}
where $P_{\text{LZ}}$ is the Landau-Zener transition probability
and $\Delta_z=(E_{+}-E_{-})|_{k_z=0,\pi/a}$ denotes the energy gap
for the  transition along the $k_{z}$ direction. Figures 5(c) and
(d) show the typical momentum distribution
$\xi_{z}(\mathbf{k_{\|}})$, where each point of maximum transfer
in the $k_x$-$k_y$ plane corresponds to a nodal point and all the
points form a nodal loop, corresponding to the cases in Figs.
\ref{probe-NL}(a) and \ref{probe-NL}(b), respectively. The peaks
in $\xi_{z}(\mathbf{k_{\|}})$ are sharp as the transition
probability in a Landau-Zener event increases exponentially as the
energy gap decreases, such that the momentum distribution of the
atomic transfer fraction with its maximum positions can well
reveal the shape of an NL.

\section{Topological properties of the simulated nodal loop states}

In this section, we precede to investigate the topological
properties of the simulated NL  states and present practical
methods for their experimental detection in this cold atom system.
We first show the quantized Berry phase characterizing the NL
states in the bulk and then consider the related non-trivial
surface states.

\subsection{Quantized Berry phase: $\mathbb{Z}_2$ topological invariant}

As analysed in the previous section, Hamiltonian (\ref{HK}) preserves the combined $\PP\TT$ symmetry.
This guarantees the symmetry-protected topological stability of
the $\PP\TT$-invariant NLs in three-dimensional  Brillouin zone
and a quantized Berry phase $\gamma$ in units of $\pi$ that characterizes its
topological protection, even in the presence of $\PP$ and $\TT$
breaking perturbation $\mathcal{H}_P$. If a closed loop in the momentum space is pierced by the NL, one has $\gamma=1$ (i.e., $\pi$ Berry phase),  otherwise $\gamma=0$, which  represents a $\mathbb{Z}_2$-type invariant. The same invariant can be written if the loop is chosen parallel to $k_z$, i.e., with fixed $\mathbf{k_{\|}}$, since it is closed at $k_z=\pm\pi/a$ due to periodic boundary conditions. Thus the topological invariant of the NL states may be evaluated as
\begin{equation}\label{BP}
\gamma (\mathbf{k_{\|}})=-\frac{i}{\pi}\sum_{E_n< E_{\text{F}}} \int_{-\pi/a}^{\pi/a} \langle u_{n}({\bf k})|\partial_{k_z}|u_{n}({\bf k})\rangle  dk_z,
\end{equation}
where the sum is over the filled Bloch eigenstates $|u_n({\bf
k})\rangle$ of Hamiltonian (\ref{HK}) with the Fermi energy $E_F$.
Equation (\ref{BP}) indicates that the topological properties of
the system can be parameterized by $\mathbf{k_{\|}}$ as the
nontrivial Zak phase for an effective gapped one-dimensional
system. Using the Bloch Hamiltonian (\ref{HK}) with given
$\mathbf{k_{\|}}$, we verify that $\gamma=1$ for
$\mathbf{k_{\|}}$ inside the projected NL while $\gamma=0$
outside, as shown in Fig. 6(a).

For fermionic atoms in this optical lattice, one can directly
probe $\gamma (\mathbf{k_{\|}})$ by measuring the  Bloch wave
function $c_{n\sigma}(\textbf{k})$ from
$|u_{n}(\textbf{k})\rangle=c_{n\uparrow}(\textbf{k})\mid\uparrow\rangle+c_{n\downarrow}(\textbf{k})\mid\downarrow\rangle$
with time-of-flight imaging \cite{Deng2014}. One first map out the
atomic momentum distribution
$\rho_{n\sigma}(\textbf{k})=|c_{n\sigma}(\textbf{k})|^{2}$ for the
filled band using the conventional time-of-flight imaging. One
then may measure the phase information of
$c_{n\sigma}(\textbf{k})$ by introducing a $\pi/2$ rotation
between the two spin states with an impulsive pulse light before
the flight of atoms, which induces the transition
$c_{n\uparrow}(\textbf{k})\rightarrow[c_{n\uparrow}(\textbf{k})+c_{n\downarrow}(\textbf{k})]/\sqrt{2}$
and
$c_{n\downarrow}(\textbf{k})\rightarrow[c_{n\uparrow}(\textbf{k})-c_{n\downarrow}(\textbf{k})]/\sqrt{2}$.
With this pulse, the difference between
$|c_{\uparrow}(\textbf{k})\pm c_{\downarrow}(\textbf{k})|^{2}/2$
measured through time-of-flight imaging gives the real part of the
interference terms
$\text{Re}[c^*_{n\uparrow}(\textbf{k})c_{n\downarrow}(\textbf{k})]$.
The imaginary part
$\text{Im}[c^*_{n\uparrow}(\textbf{k})c_{n\downarrow}(\textbf{k})]$
can be obtained by the same way with a different rotation. The
measurement of the population and interference terms determines
the Bloch wave function up to an arbitrary overall phase
$c_{n\sigma}(\textbf{k})\rightarrow
c_{n\sigma}(\textbf{k})e^{i\chi(\textbf{k})}$, where
$\chi(\textbf{k})$ in general depends on $\textbf{k}$ instead
of the spin index. The arbitrary $\textbf{k}$-dependent phase
poses an obstacle to measure the topological invariant
\cite{Deng2014}. To overcome this difficulty, we use a
gauge-invariant method to calculate the Berry phase \cite{BPComp}
\begin{equation} \label{BPv2}
\gamma (\mathbf{k_{\|}})= \frac{1}{\pi}\sum_{j=0}^{N_{j}-1} \text{Arg}[\text{det}\langle u_{n}(\mathbf{k_{\|}},k^z_{j})|u_{n'}(\mathbf{k_{\|}},k^z_{j+1})\rangle],
\end{equation}
where $k_z$ in the BZ is discretized into small $N_j$ intervals
with $k^z_{j} = -\pi/a+2j\pi/N_ja$, the overlap  phase
$\text{Arg}[\bullet]=\text{Imag}\{\ln[\bullet]\}$, and the
determinant is that of a matrix formed by allowing $n$ and $n'$ to
run over filled Bloch eigenstates.

\begin{figure}[tbph]
\includegraphics[width=8.5cm]{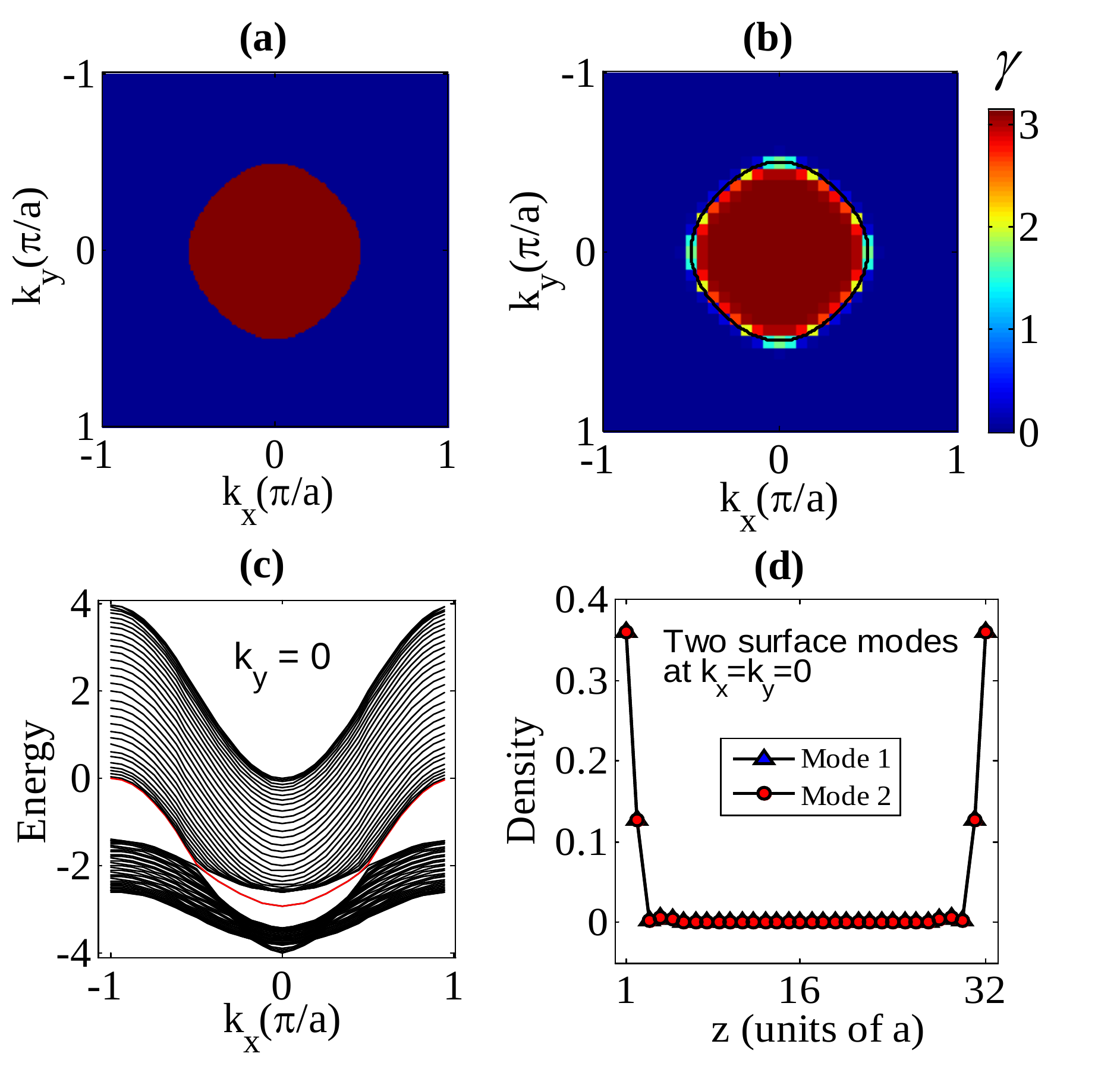}
\caption{(Color online) (a) The Berry phase $\gamma(k_x,k_y)$
calculated by using the Bloch Hamiltonian (\ref{HK}). (b)  The
simulated measurement of Berry phase $\gamma(k_x,k_y)$ with the
lattice size $32\times32\times32$ and an additional harmonic trap
with the parameter $\nu=0.001$ (see the text) by using Eq.
(\ref{BPv2}). The black solid line denotes the nodal loop in this
case. (c) The energy spectrum with respect to $k_x$ for fixed
$k_y=0$ and open boundary condition along the $z$ axis, and the
surface states inside the gap. (d) The density distribution of two
surface modes at $k_x=k_y=0$. Other parameters in (a-d) are
$t_{\uparrow}=1$ as the energy unit, $t_{\downarrow}=t_{\rm
so}=0.3$, and $m_z=2.0$.\label{Berry-edge}}
\end{figure}

To demonstrate that the method is feasible in a realistic
experiment, we numerically simulate  the proposed detection of the
Berry phase $\gamma (\mathbf{k_{\|}})$ with finite lattice system
and an additional harmonic trap
\begin{equation} \hat{H}_{\text{trap}}=\frac{1}{2}m_a\omega^2\sum_{\textbf{i},\sigma}d^2_{\textbf{i}}\hat{a}^{\dag}_{\textbf{i},\sigma}\hat{a}_{\textbf{i},\sigma},
\end{equation}
where $\omega$ is the trap frequency and $d_{\textbf{i}}$ is the
distance from the center of the  trap to the lattice site
$\textbf{i}$. We can use $\nu=m_a\omega^2a^2/2 t_{\uparrow}$ to
parametrize the influence of this trapping potential. For a
typical experiment with $a\approx400$ nm and
$t_{\uparrow}\approx1$ kHz, $\mu$ is on the order of $10^{-3}$ for
$^6$Li or $^{40}$Ka atoms in a trap with
$\omega\approx2\pi\times50$ Hz. In numerical simulations, we
perform the spin rotation and obtain the momentum distribution
under different spin basis by diagonalizing the real-space
Hamiltonian on a finite lattice and using a Fourier transformation
\cite{Deng2014}. The numerical result of $\gamma
(\mathbf{k_{\|}})$ for finite lattice $32\times32\times32$ with a
weak harmonic trap ($\nu=0.001$) and typical parameters is shown
in Fig. \ref{Berry-edge}(b). Compared with the result shown in
Fig. \ref{Berry-edge}(a), in this case the $\mathbf{k_{\|}}$
regime with $\gamma=1$ slightly shrink from that of the ideal NL
[black solid line in Fig. \ref{Berry-edge}(b)] and the sharp
boundary between $\gamma=0$ and $\gamma=1$ becomes relatively
smooth, which are due to the trapping potential and the finite
size effects.

\subsection{Protected surface states}

According the the bulk-edge correspondence, the topological NL
state with the Berry  phase $\gamma (\mathbf{k_{\|}})$ is related
to edges states at the end of the one-dimensional system with a
fixed $\mathbf{k_{\|}}$ \cite{NodalTheo1,BPComp,JR}. Hence, for a
fixed $\mathbf{k_{\|}}$ and two surfaces (upper and lower
surfaces) along the $z$ axis, two in-gap states appear at in the
surface Brillouin zone when $\gamma (\mathbf{k_{\|}})\neq0$. This
indicates that surface states appear for all $\gamma
(\mathbf{k_{\|}})$ inside the area enclosed by the projected NL on the surface plane. With the $f_0(\mathbf{k})\sigma_0$
term, the surface states acquire a dispersion proportional to
$f_0(k_x,k_y,0)$. In Fig. \ref{Berry-edge}(c), we
plot the energy spectrum with respect to $k_x$ for fixed $k_y=0$
and open boundary condition along the $z$ axis (with 32 lattice
sites), which shows the surface states inside the gap denoted by
red solid lines connecting the two band-touching points. The
density distribution of the two corresponding surface modes at
$k_x=k_y=0$ is shown in Fig. \ref{Berry-edge}(d). The surface
states for other $k_y$ are similar, and they merge into the bulk
bands when the system become a trivial insulator. For this system
in phase II.a, the surface modes perpendicular to the $z$
direction with fixed $\mathbf{k_{\|}}$ can be described by the
effective Hamiltonian
$H_z(\mathbf{k_{\|}})=iv_z\sigma_2\partial_z+g_z(\mathbf{k_{\|}})\sigma_3+g_0(\mathbf{k_{\|}})\sigma_0$
up to the linear order in $k_z\rightarrow-i\partial_z$, where
$g_z(\mathbf{k_{\|}})=m_z-\alpha_+-\alpha_{-}[\cos(k_xa)+\cos(k_ya)]$
and
$g_0(\mathbf{k_{\|}})=-\alpha_{-}-\alpha_{+}[\cos(k_xa)+\cos(k_ya)]$.

In the optical lattice, the surface states may be washed out by
the smooth harmonic potential and become undistinguishable from
the bulk states. To circumvent this problem, one can use a steep
confining potential or cut the atomic hopping along the $z$ axis
by locally tuning the effective Rabi frequency of the Raman
lasers. Under this condition, the protected surface states can be
probed through Bragg spectroscopy
\cite{Duan2014,zhu2007,Goldman2012}. One could shine another two
laser beams at a certain angle to induce a specifically tuned
Raman transition from an occupied spin state to an unoccupied
hyperfine level and focus them near the surface of the
three-dimensional atomic cloud \cite{Goldman2012}. When the
momentum and energy conservation conditions are satisfied, the
atomic transition rate is peaked and can be measured. By scanning
the Raman frequency difference, the surface energy-momentum
dispersion relation can be mapped out from these Bragg signals
\cite{zhu2007,Goldman2012}.

\section{conclusion}

In summary, we have presented a topological classification of
NLs in systems with the $\PP\TT$ symmetry. In three-dimensional
momentum space, only $\PP\TT$-invariant NLs are topologically
protected with a $\mathbb{Z}_2$ classification. Motivated by this
observation, we have proposed a realistic experimental scheme to
realize $\PP\TT$-invariant topological NL states with cold atoms
in a three-dimensional optical lattice, which have tunable
loop-shaped Fermi lines with two-fold degeneracy in the bulk
spectrum and non-trivial surface states. The NL states are
actually protected by the combined $\PP\TT$ symmetry even in the
absence of both $\PP$ and $\TT$ symmetries, and are
characterized by a quantized Berry phase (a $\mathbb{Z}_2$-type
invariant). We have also shown that (i) the characteristic NLs can be detected by measuring the atomic transfer fractions in
a Bloch-Zener oscillation; (ii) the topological charge can be
measured based on the time-of-flight imaging; and (iii) the
surface states can be probed through Bragg spectroscopy. The
experimental realization and detection of $\PP\TT$-symmetry
protected NL states in cold atom systems will be regarded as an
important advance in the field of quantum simulation, paving the
way for exploring exotic $\PP\TT$-invariant topological physics.

\section{Acknowledgments}

We thank M. Gong for useful discussions. This work was supported
by the NSFC (Grant No. 11474153), the PCSIRT (Grant No. IRT1243),
the SKPBR of China (Grant No. 2013CB921804), the SRFYTSCNU (Grant No. 15KJ16), and the RGC of Hong
Kong (HKU173051/14P and HKU173055/15P).


\end{document}